\documentclass[aps,prl,10pt,superscriptaddress,twocolumn,shortbibliography,showkeys,serif]{revtex4-2}

\usepackage{graphicx}
\usepackage{xcolor}
\usepackage{hyperref}
\usepackage{amsfonts, amsmath, amssymb}
\usepackage{times}
\usepackage{dcolumn}
\usepackage{multirow}
\usepackage{enumitem}
\usepackage{booktabs}
\usepackage{subcaption}
\usepackage[sort&compress]{natbib}

\hypersetup{colorlinks=true, linkcolor=blue, urlcolor=blue, citecolor=blue}

\graphicspath{{figs/}}

\begin{document}

\title{A New $\sim 5\sigma$ Tension at Characteristic Redshift from DESI-DR1 BAO and DES-SN5YR Observations}

\author{Purba Mukherjee}
\email{pdf.pmukherjee@jmi.ac.in}
\affiliation{Centre for Theoretical Physics, Jamia Millia Islamia, New Delhi-110025, India}%

\author{Anjan A Sen}
\email{aasen@jmi.ac.in}
\affiliation{Centre for Theoretical Physics, Jamia Millia Islamia, New Delhi-110025, India}%

\date{\today}

\begin{abstract}
We perform a model-independent reconstruction of the angular diameter distance ($D_{A}$) using the Multi-Task Gaussian Process (MTGP) framework with DESI-DR1 BAO and DES-SN5YR datasets. We calibrate the comoving sound horizon at the baryon drag epoch $r_d$ to the Planck best-fit value, ensuring consistency with early-universe physics. With the reconstructed $D_A$ at two key redshifts, $z\sim 1.63$ (where $D_{A}^{\prime} =0$) and at $z\sim 0.512$ (where $D_{A}^{\prime} = D_{A}$), we derive the expansion rate of the Universe $H(z)$ at these redshifts. Our findings reveal that at $z\sim 1.63$, the $H(z)$ is fully consistent with the Planck-2018 $\Lambda$CDM prediction, confirming no new physics at that redshift. However, at $z \sim 0.512$, the derived $H(z)$ shows a more than $5\sigma$ discrepancy with the Planck-2018 $\Lambda$CDM prediction, suggesting a possible breakdown of the $\Lambda$CDM model as constrained by Planck-2018 at this lower redshift. This emerging $\sim 5\sigma$ tension at $z\sim 0.512$, distinct from the existing ``Hubble Tension'', may signal the first strong evidence for new physics at low redshifts.  
\end{abstract}


\maketitle


In recent decades, cosmology has witnessed remarkable progress, largely driven by high-precision observations from the Cosmic Microwave Background (CMB) \cite{Planck:2018vyg, ACT:2020gnv, Tristram:2023haj}, type Ia supernovae (SNIa) \cite{Blanchard:2022xkk, Peebles:2024txt}, large-scale structure (LSS), and galaxy surveys \cite{BOSS:2014hhw, BOSS:2016wmc, eBOSS:2020yzd, Addison:2017fdm}. The standard model of cosmology, $\Lambda$CDM, describes the universe as being predominantly composed of cold dark matter (CDM) and dark energy (DE), with the latter represented by a cosmological constant ($\Lambda$) that drives the late-time accelerated expansion of the universe \cite{Weinberg:1988cp, Sahni:1999gb}.

Despite its success, the $\Lambda$CDM model faces several challenges, as recent observations have revealed notable inconsistencies \cite{Hazra:2013dsx, Verde:2019ivm, Perivolaropoulos:2021jda,  Efstathiou:2024dvn}. One of the most debated issues is the Hubble tension \cite{DiValentino:2020zio, Riess:2021jrx}, where local measurements of the Hubble constant ($H_0$) differ by more than $5\sigma$ from values inferred using CMB data. Another discrepancy, known as the $\sigma_8$ tension \cite{DiValentino:2020vvd}, refers to a $\approx 2.5\sigma$ mismatch between the CMB-predicted vs observed clustering of matter, measured by galaxy surveys \cite{DES:2021wwk,Heymans:2020gsg,Li:2023tui}. Furthermore, deep-space observations from the James Webb Space Telescope (JWST) have uncovered unexpectedly massive and bright galaxies at redshifts $z \gtrsim 7$ \cite{Boylan-Kolchin:2022kae, Labbe:2022ahb}, posing additional challenges to the concordance framework. These discrepancies indicate potential gaps in our understanding of cosmic evolution and structure formation, suggesting the need for extensions or modifications beyond the conventional six-parameter $\Lambda$CDM model \cite{DiValentino:2021izs, Abdalla:2022yfr}.  

Recent baryon acoustic oscillation (BAO) measurements from the Dark Energy Spectroscopic Instrument (DESI) indicate that the equation of state (EoS) of dark energy may evolve over cosmic time, with hints of an early phantom-like behavior \cite{DESI:2024mwx}. While DESI observations remain consistent with the standard $\Lambda$CDM model, extending the analysis to the Chevallier-Polarski-Linder (CPL) parameterization reveals deviations from $w = -1$. When combined with CMB and the recent five-year SNIa data from the Dark Energy Survey Supernova Program (DES-SN5YR) \cite{DES:2024tys}, the CPLCDM model offers a statistically better fit, excluding $\Lambda$CDM at a confidence level of more than $3.9\sigma$ \cite{DESI:2024mwx}. These findings have reignited interest in exploring alternative scenarios to probe potential new physics in the cosmic dark sector \cite{Colgain:2024mtg, liddle_desi, Park:2024jns, Wolf:2025jlc, wang_desi, Calderon:2024uwn, DESI:2024kob, Gialamas:2024lyw, Jiang:2024xnu, Dinda:2024ktd, Bhattacharya:2024hep, RoyChoudhury:2024wri, Lima:2025yza, Keeley:2025stf, Choudhury:2025bnx, Tiwari:2024gzo, Barua:2024gei, Ferrari:2025egk, Mukherjee:2025myk, Huang:2025som, Bansal:2025ipo, Sousa-Neto:2025gpj, Alfano:2025gie, Sakr:2025daj, Akarsu:2025gwi, Hossain:2025grx, Shah:2024gfu, Heckman:2024apk, Montani:2024ntj, Colgain:2024ksa}.

In this paper, we reconstruct the angular diameter distance, $D_A$, and its derivative, $D_A'$, in a cosmological model-independent framework. The angular diameter distance $D_A(z)$ is defined as:
\begin{equation}
D_{A}(z) = \frac{c}{(1+z)}\int^{z}_{0} \frac{1}{H(z)} dz.
\end{equation}
Here $c$ is the speed of light. In this definition, the part involving the integration of the inverse of the Hubble parameter $H(z)$ is related to the comoving distance. The presence of the term $(1+z)$ in the denominator is related to the evolution of the distance between two objects (e.g two galaxies) due to the expanding Universe. Although comoving distances increase with redshift, the separation of two objects in the Universe decreases with redshift in an expanding universe. Due to this, the angular diameter distance $D_{A}$ has a maximum in the past, which is a distinctive feature, not present in other observables related to distance, e.g. luminosity distance $D_{L}$. This feature allows us to measure the $H(z)$ at redshift $z=z_{1}$ where $D_{A}$ is maximum, directly from the $D_{A}$ itself. This is because the derivative of $D_{A}$ is given by:
\begin{equation}
    D_{A}^{\prime} = \frac{c}{H(z)(1+z)} - \frac{D_{A}}{(1+z)}.
    \end{equation}

\noindent
Here ``{\it prime}" represents derivative with respect to redshift $z$. Now if at $z=z_{1}$, $D_{A}$ is maximum and hence $D_{A}^{\prime} = 0$, then
\begin{equation}  
H(z_1) = \frac{c}{D_A(z_1)} \, ,  
\end{equation} 

\noindent
This offers a robust model-agnostic determination of $H(z=z_{1})$ directly from the $D_{A}$ observational data.

We can do further. Note that at $z=0$, $D_{A} = 0$, whereas $D_{A}^{\prime} = \frac{c}{H_{0}}$. As $z$ increases $D_{A}$ increases and attains a maximum at $z=z_{1}$ whereas $D_{A}^{\prime}$ decreases and goes to zero at $z=z_{1}$. Hence $D_{A}$ and $D_{A}^{\prime}$ should cross at some redshift $z_{2} < z_{1}$, where $D_{A}(z=z_{2}) = D_{A}^{\prime}(z=z_{2})$. This also allows to measure $H(z)$ from $D_{A}$ at $z=z_{2}$ through the relation:
\begin{equation}  
H(z_2) = \frac{c}{D_A(z_2)} (z_2 + 2)^{-1}. 
\end{equation}  

Thus, the characteristic redshifts $z_1$ and $z_2$ are defined by the following conditions:  
\begin{itemize}[left=0pt]
\item At $z \equiv z_1$: $D_{A}^\prime(z_1) = 0$, leading to  $H(z_1) = \frac{c}{D_A(z_1)}.$
\item At $z \equiv z_2$: $D_{A}^\prime(z_2) = D_A(z_2)$, yielding  $H(z_2) = \frac{c}{D_A(z_2)} (z_2 + 2)^{-1}.$
\end{itemize}

This gives a robust model-agnostic measurements of the Hubble parameter $H(z)$ at these two redshifts $z_{1}$ and $z_{2}$ from observational data related to angular diameter distance and subsequently allows us to test whether $H(z_{1})$ and $H(z_{2})$ align with Planck-2018 measurements of $H(z)$ at these two redshifts assuming a concordance $\Lambda$CDM model. Any inconsistencies in $H(z_1)$ or $H(z_2)$ could signal new physics or systematics in observational data.


\noindent We use the following datasets in our analysis:
\begin{itemize}[left=0pt]
    \item[-] Transverse comoving distance $D_M(z)/r_d$, and comoving Hubble distance $D_H(z)/r_d$ measurements from DESI-BAO DR1 \cite{DESI:2024mwx}, covering the redshift range $0.1 < z < 4.2$. Here, $r_d$ is the sound horizon at the drag epoch. Henceforth, we denote this dataset as `DESI'.
    \item[-] DES-SN5YR \cite{DES:2024tys, DES:2024upw, the_des_sn_working_group_2024_12720778} sample of 1635 photometrically-classified SNIa, in the redshifts range $0.1 < z < 1.3$, complemented by 194 low-$z$ SNIa (overlapping with Pantheon+ \cite{Scolnic:2021amr} compilation) in the range $0.025 < z < 0.1$. Hereafter, we refer to this dataset as `DESY5'.
\end{itemize}

We aim to reconstruct the angular diameter distance $D_A(z)$ and its derivative $D_{A}^\prime(z)$ in a model-agnostic manner using Multitask Gaussian process (MTGP) regression \cite{mukherjee:2025, Mukherjee:2024ryz, Dinda:2024ktd, Haridasu:2018gqm,Perenon:2021uom}, utilizing the latest BAO and SNIa data sets. The BAO observables are $D_M(z)/r_d$ and $D_H(z)/r_d$ where $D_{M}(z) = c\int^{z}_{0} \frac{dz}{H(z)}$ is the comoving distance and $D_{H} = \frac{c}{H(z)}$. $r_{d}$ is the comoving sound horizon at the baryon drag epoch.


\squeezetable
\begin{table*}
    \centering
    \renewcommand{\arraystretch}{1.3} \setlength{\tabcolsep}{7pt}
    \begin{tabular}{lccccccc}
        \hline \toprule
         Model & $H_0$ & $z_1$ & $D_A(z_1)$ & $H(z_1)$ & $z_2$ & $D_A(z_2)$ & $H(z_2)$ \\
        \hline \midrule 
        MTGP SE & $67.20 \pm 0.35$ & \multirow{2}{*}{$1.646 \pm 0.040$} & $1772.83 \pm 12.54 $ & $ 169.11 \pm 1.20$ & \multirow{2}{*}{$0.512 \pm 0.002$} & $1297.17 \pm 4.37 $ & $ 92.01 \pm 0.31$ \\
        P18 $\Lambda$CDM & $67.36 \pm 0.54$ & & $1794.24 \pm 4.62 $ & $ 172.38 \pm 3.52$ & & $1317.65 \pm 7.63 $ & $ 89.74 \pm 0.29$ \\
        \midrule
        Tension & $0.25\sigma$ & & $1.6\sigma$ & ${0.88\sigma}$ &  & $2.33\sigma$ & $\boldsymbol{5.39\sigma}$ \\
        \midrule
        MTGP M72 & $67.17 \pm 0.39$ & \multirow{2}{*}{$1.625 \pm 0.048$} & $1774.11 \pm 12.36$ & $168.98 \pm 1.18$ & \multirow{2}{*}{$0.512 \pm 0.002$} & $1296.71 \pm 4.37$ & $92.03 \pm 0.31$ \\
        P18 $\Lambda$CDM & $67.36 \pm 0.54$ & & $1794.64 \pm 4.65$ & $170.32 \pm 4.20$ & & $1318.19 \pm 7.53$ & $89.75 \pm 0.28$ \\
        \midrule
        Tension & $0.29\sigma$ & & $1.56\sigma$ & $0.31\sigma$ & & $2.47\sigma$ & $\boldsymbol{ 5.42\sigma}$ \\
        \midrule
        MTGP M92 & $67.17 \pm 0.37$ & \multirow{2}{*}{$ 1.638 \pm 0.043$} & $1775.33 \pm 12.39$ & $168.87 \pm 1.18$ & \multirow{2}{*}{$0.512 \pm 0.002$} & $1298.20 \pm 4.35 $ & $ 91.93 \pm 0.30 $ \\
        P18 $\Lambda$CDM & $67.36 \pm 0.54$ & & $1794.46 \pm 4.63$ & $ 171.38 \pm 3.80 $ & & $1317.82 \pm 7.57 $ & $89.73 \pm 0.29$\\
        \midrule
        Tension & $0.29\sigma$ & & $1.45\sigma$ & $0.63\sigma$ & & $2.25\sigma$ & $\boldsymbol{5.23\sigma}$\\
        \midrule
        MTGP RQ & $67.31 \pm 0.34$ & \multirow{2}{*}{$1.626 \pm 0.032$} & $1770.84 \pm 11.72$ & $169.29 \pm 1.12$ & \multirow{2}{*}{$0.513 \pm 0.002$} & $1299.42 \pm 4.27$ & $91.84 \pm 0.30$ \\
        P18 $\Lambda$CDM & $67.36 \pm 0.54$ &  & $1794.73 \pm 4.62 $ & $ 170.33 \pm 2.84$ &  & $1319.60 \pm 7.59 $ & $ 89.80 \pm 0.29$\\
        \midrule
        Tension & $0.08\sigma$ & & $1.89\sigma$ & $0.34\sigma$ &  &  $2.32\sigma$ & $\boldsymbol{5.1\sigma}$\\
        \midrule
        MTGP Linear & $67.25 \pm 0.39$ & \multirow{2}{*}{$1.603 \pm 0.04$} & $1774.41 \pm 12.03$ & $168.95 \pm 1.15$ & \multirow{2}{*}{$0.512 \pm 0.002$} & $1297.42 \pm 4.36$ & $91.99 \pm 0.31$ \\
        $P18 \Lambda$CDM & $67.36 \pm 0.54$ & & $1794.85 \pm 4.67$ & $168.41 \pm 3.50$ & & $1317.53 \pm 7.53$ & $89.73 \pm 0.28$ \\
        \midrule
        Tension & $0.13\sigma$ & & $1.51\sigma$ & $0.15\sigma$ &  & $2.31\sigma$ & $\boldsymbol{5.41\sigma}$ \\
        \bottomrule \hline
    \end{tabular}
     \caption{Table showing the result from reconstruction at characteristic redshifts $z_1$ and $z_2$. Here $H_0$ and $H(z)$ are in units of km Mpc$^{-1}$ s$^{-1}$, while $D_A(z)$ and $r_d$ are in units of Mpc.} \label{tab:result}
\end{table*}

 The BAO data can provide direct data-driven constraints on $H_0 r_d$, meaning that the comoving sound horizon at the baryon drag epoch \cite{Eisenstein:1997ik}, $r_d$, must be calibrated to determine $H_0$ and infer $H(z)$, or vice versa. The SNIa data, on the other hand, contain information about $E(z)$ through the luminosity distance $D_{L}$ but require calibration of the absolute magnitude $M_B$, which is degenerate with $H_0$ and is treated as an astrophysical nuisance parameter in cosmological analysis. Assuming there are no underlying systematics between BAO and SNIa observations, we undertake this exercise via an early-universe calibration for $r_d$. We consider $r_d$ is determined by early-universe physics, adopting the prior $r_d = 147.09 \pm 0.27$ Mpc \cite{Planck:2018vyg}. This allows us to marginalize over $M_B$ and infer the evolutionary profile in a cosmological model-agnostic way (see Mukherjee \& Sen\cite{Mukherjee:2024ryz} for methodological details). Our approach follows along the lines of \cite{camilleri2024darkenergysurveysupernova}, which employs third, fourth, and fifth-order cosmography to reconstruct the inverse distance ladder.

In what follows, we work out the MTGP reconstruction of $D_A$ and $D_{A}^\prime$ using the DESI+DESY5 datasets to explore potential deviations from Planck $\Lambda$CDM. Our key motivation is to test departures from Planck $\Lambda$CDM at the {\it characteristic redshifts} $z_1$ and $z_2$, highlighting interesting features in the evolution of the universe. For an exhaustive analysis, we take into account different choices for the GPR kernel - namely Mat\'ern 7/2 (M72), Mat\'ern 9/2 (M92), Squared Exponential (SE) and Rational Quadratic (RQ) covariance functions. One can refer to Mukherjee \& Sen\cite{Mukherjee:2024ryz} for mathematical details on the MTGP framework and possible kernel choices. Additionally, we evaluate the impact of imposing a zero-mean function versus a linear-in-redshift mean function to assess the robustness of our results. The kernel hyperparameters, along with the parameters governing the mean function, are constrained using the \href{https://github.com/dfm/tinygp.git}{\textsc{tinygp}} \cite{foreman_mackey_2024_10463641} module, which implements Bayesian MCMC analysis with \href{https://github.com/jax-ml/jax.git}{\textsc{jax}} \cite{jax2018github} and \href{https://github.com/pyro-ppl/numpyro.git}{\textsc{numpyro}} \cite{phan2019composable, bingham2019pyro}. For this purpose, we assume uniform flat priors on kernel hyperparameters and the mean function parameters. We compute the Hubble parameter $H(z)$ and angular diameter distances $D_A(z)$, at these characteristic redshifts $z_1$ and $z_2$, directly from the Planck (2018) TTTEEE+lowE+lensing likelihood chains and derive the Gaussian tension metric between the MTGP reconstructed values vs Planck 2018 predictions. 

\begin{figure}
    \centering
    \includegraphics[width=0.95\linewidth]{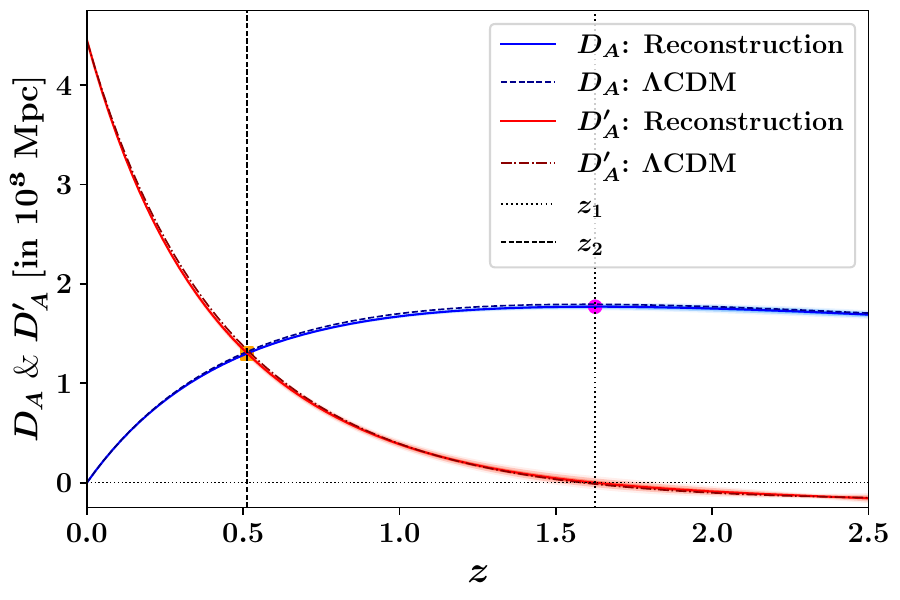}
    \caption{Reconstructed evolution for $D_A(z)$ and $D_A'(z)$ as a function of redshift, with RQ kernel. The shaded regions correspond to the 1$\sigma$ and 2$\sigma$ confidence levels. The two characteristic redshifts $z_1$ (where $D_A'(z_1)=0$) and $z_2$ ($D_A'(z_2)=D_A(z_2)$) are marked with circle and square markers. The Planck $\Lambda$CDM best-fit curves for $D_A(z)$ and $D_A'(z)$ are shown in dashed lines.}
    \label{fig:DA_vs_z}
\end{figure}

The results of our analysis are presented in Table \ref{tab:result}. Figure \ref{fig:DA_vs_z}, shows the evolution of reconstructed $D_A(z)$ and $D_A^\prime(z)$ as a function of redshift. The shaded regions correspond to the 1$\sigma$ and 2$\sigma$ confidence levels. The two characteristic redshifts $z_1$ and $z_2$ are marked with circle and square markers. For comparison, we also plot the $D_A$ and $D_A'$ from Planck $\Lambda$CDM best-fit. The {\it early-universe calibration} is done to assess the consistency of the reconstructed expansion history with the Planck 2018 $\Lambda$CDM model. We set the sound horizon scale to the Planck best-fit value $r_d = 147.09 \pm 0.27$ Mpc, ensuring that our reconstruction is anchored to early-universe physics as inferred by Planck. Despite this $-$ which, under the assumption that $\Lambda$CDM accurately describes the universe, should ensure compatibility with Planck baseline predictions $-$ we still observe significant deviations in the evolutionary profile of the expansion history, particularly in the angular diameter distance $D_A(z)$ and Hubble parameter $H(z)$. The statistical tension, quantified in standard deviations ($\sigma$), is particularly strong for $H(z_2)$, reaching up to $\sim 5\sigma$, and remains non-negligible for $D_A(z_1)$ and $D_A(z_2)$. This suggests that while setting $r_d$ to Planck values ensures agreement at the calibration point, the reconstructed evolution at intermediate redshifts prefers a different trajectory from that of Planck $\Lambda$CDM. This deviation indicates that the data is favoring an expansion history that does not fully conform to the standard $\Lambda$CDM expectations, pointing towards potential new physics or systematic differences in the observational sector. \\


\begin{figure}
    \centering
    \includegraphics[width=0.95\linewidth]{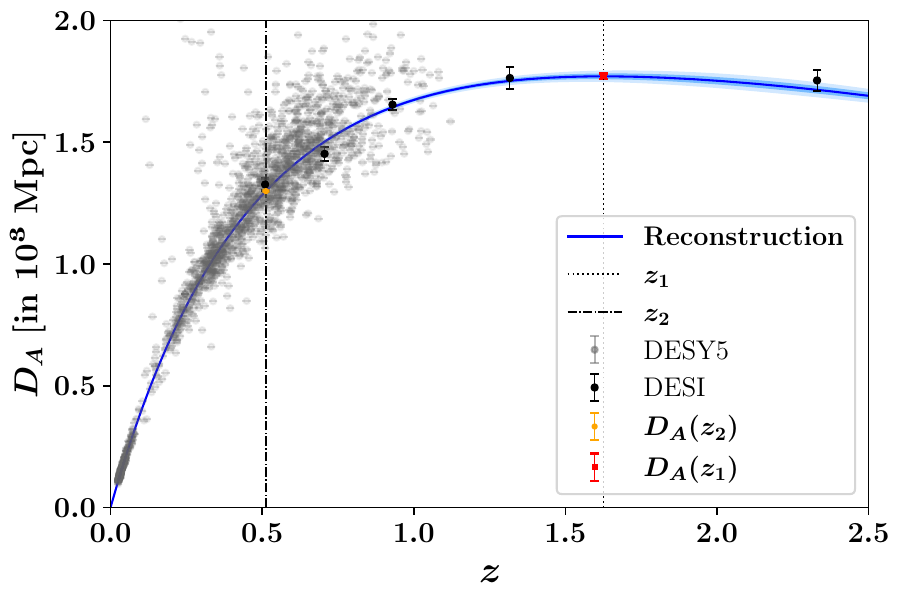}
    \caption{Reconstruction of $D_A$ vs $z$. The DESY5 and DESI data points are shown for comparison. The reconstructed vales of $D_A(z_1)$ and $D_A(z_2)$ are marked for illustration.}
    \label{fig:Hz_vs_z}
\end{figure}


\begin{figure}
    \centering
    \includegraphics[width=0.95\linewidth]{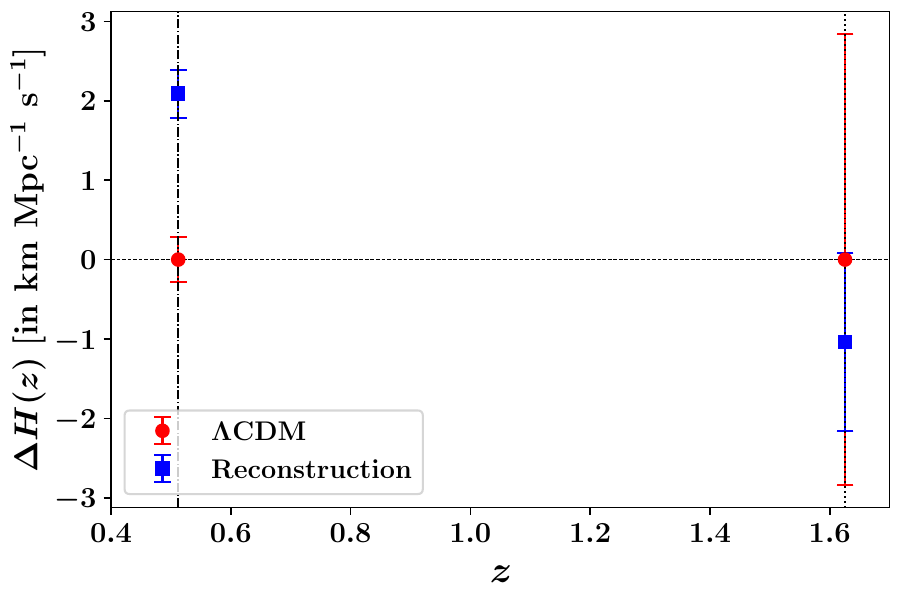}
    \caption{Residuals $\Delta H(z_1) \equiv H(z_1)^{\text{Reconstruction}} - H(z_1)^{\text{$\Lambda$CDM}}$ and $\Delta H(z_2) \equiv H(z_2)^{\text{Reconstruction}} - H(z_2)^{\text{$\Lambda$CDM}}$, assuming the RQ kernel for reconstruction.}
    \label{fign:Hz_vs_z}
\end{figure}

\noindent Our findings have been summarized as follows$-$
\begin{enumerate}[left=0pt]
    \item {\bf Robustness of the reconstructed values :} The reconstructed values for $D_A$ and $H$ at characteristic redshifts $z_1$ and $z_2$, as presented in Table \ref{tab:result}, remain robust under different choices of kernels and mean functions. The fact that our findings do not significantly change with different covariance structures and prior assumptions indicates that the inferred constraints are credible and resilient against methodological variations. 
    \item {\bf Consistency of $H_0$ with Planck and DESI+DESY5 constraints :} Our constraints on the Hubble constant $H_0$ are fully compatible with both the Planck $\Lambda$CDM ($H_0^{\rm P18} = 67.36 \pm 0.54$ km Mpc$^{-1}$ s$^{-1}$ \cite{Planck:2018vyg}) and the DESI+DESY5 ($H_0 = 67.19^{+0.66}_{-0.64}$ km Mpc$^{-1}$ s$^{-1}$ \cite{camilleri2024darkenergysurveysupernova}) constraints. On fixing the comoving sound horizon scale to the Planck best-fit value, $r_d = 147.09$ Mpc, we find that the inferred value of $H_0$ remains consistent with Planck's baseline predictions. This reinforces that our early-universe calibration method aligns well with the expectations from the concordance model, supporting the robustness of our approach. 
    \item {\bf No significant tension at $z_1$ (where $D_{A}^\prime = 0$) :} At the characteristic redshift $z_1$, where the derivative of the angular diameter distance vanishes, our reconstructed constraints show no significant deviations in either $D_A(z_1)$ or $H(z_1)$. There is a 2-2.5\% uncertainty associated with determining $z_1$, and it varies between different kernel choices. But our results for $D_A(z_1)$ and $H(z_1)$ are quite robust, showing no abrupt change in the evolutionary history or dark energy behaviour around that redshift $z_1 \sim 1.6 $ (for example, in Ref. \cite{Akarsu_2023, akarsu2023lambdarmscdmmodelpromising, Akarsu:2025gwi}).
    \item {\bf Strong $>5\sigma$ tension at $z_2$ (where $D_A = D_{A}^\prime$) :} A significant discrepancy is observed at the characteristic redshift $z_2$, where the angular diameter distance equals to its derivative. The tension at this redshift exceeds $5\sigma$, marking a strong deviation from the Planck $\Lambda$CDM model. 
    Moreover, $z_2$ is much more precisely determined with less than $1\%$ uncertainty.

    \item {\bf Possible hints for new physics at low-redshift :} Given the significant tension observed at very low redshift, one possible explanation could be unknown systematic uncertainties in the DESI and DESY5 data. Interestingly, a similar deviation is also present in the inverse distance ladder analysis by DES-SN5YR Collaboration (see Fig. 1 of \cite{camilleri2024darkenergysurveysupernova}). If this 5$\sigma$ tension is not attributed to unknown systematics in the DESI or DESY5 data \cite{efstathiou2025evolvingdarkenergysupernovae, vincenzi2025comparingdessn5yrpantheonsn}, then resolving such a discrepancy through early-time modifications or pre-recombination physics appears unlikely. It possibly demands new physics at redshift around $z \sim 0.512$ \cite{Vagnozzi_2023}.

     \item We also study how far our results are affected by the DESI Luminous Red Galaxy (LRG) data at $z_{\rm eff} = 0.51$ in the redshift range $0.4 < z < 0.6$ \cite{eoin_desi, Sapone:2024ltl, wang_desi}. To this effect, we exclude this data point and redo the analysis. The $H(z)$ tension at $z_{1}$ is $\sim 0.46 \sigma$ and at $z_{2}$ is at $\sim 4.97\sigma$, showing the LRG data does not affect the tension much.

     \item{We should mention that our result for $H(z)$ at $z_{1} \sim 1.63$ is driven by the DESI data as around that redshift there is no DES-SN5YR measurement. Similarly, at the characteristic redshift $z_{2} \sim 0.512$, the $H(z)$ determination is mostly governed by the  DES-SN5YR data as there are a large number of DES-SN5YR data around that redshift as compared to the single DESI LRG1 data at $z_{\rm eff}=0.51$.   This is shown in Figure \ref{fig:Hz_vs_z}.  However, our results are not very sensitive to this DESI data, as discussed above. }

    \item Given our results for $H(z)$ at $z\sim 1.63$ and $z\sim 0.512$ and corresponding values for $H(z)$ from Planck-2018 for $\Lambda$CDM as shown in Table \ref{tab:result}, we notice that (as shown in Figure \ref{fign:Hz_vs_z}) in the redshift range $0.512<z<1.63$, the $H(z)$ decreases in a slower rate than as predicted by Planck-2018 $\Lambda$CDM. This gives a possible hint for dark energy evolution (a non-phantom one) in the redshift range $0.512<z<1.63$ unless there are abrupt changes in $H(z)$ evolution in this redshift interval.

\end{enumerate}


To conclude, we employ a novel technique using the features in the angular diameter distance $D_{A}(z)$ to obtain model-agnostic values for $H(z)$ at two characteristic redshifts $z_{1} \sim 1.63$ and $z_{2} \sim 0.512$. We use the combination of DESI-DR1 BAO and DES-SN5YR measurements and the Multi-Task Gaussian Process (MTGP) framework for this purpose. We show that at $z=z_{1}$ the $H(z)$ is fully consistent with Planck-2018 $\Lambda$CDM predictions. But at $z=z_{2}$, the $H(z)$ value obtained is at more than $5\sigma$ tension with the Planck-2018 $\Lambda$CDM prediction, showing a new tension in the expansion rate of the Universe at low redshift. These results are stable under different choices of the kernel and mean functions, confirming that the results are robust. Unless there is a large systematics in the DESI or DESY5 data, this new $\sim 5\sigma$ tension at $z=0.512$ in $H(z)$ confirms the possible breakdown of the $\Lambda$CDM model as constrained by Planck-2018. This is independent of already existing $5\sigma$ tension from SH0ES measurement. In our analysis, we use the calibration using the early Universe sound horizon scale at drag epoch $r_{d}$ given by the Planck-2018. One can similarly use the calibration using the local measurement of $H_{0}$ by SH0ES \cite{Riess:2021jrx}. This can change our results substantially. A back of the envelope calculation shows that using $H_{0}$ calibration from SH0ES instead of $r_{d}$ calibration from Planck-2018, increases the tension at $z_{1}\sim 1.63$ to around $3\sigma$ whereas the the tension at $z_{2} \sim 0.512$ increases to much higher value. But one needs to do a full analysis in this regard, which we aim to do in the near future.
\\

\noindent {\textbf{Note added:}} In a recent study by Ormondroyd \textit{et al}\cite{Ormondroyd:2025exu}, the authors performed a non-parametric, free-form reconstruction of the dark energy equation of state. They found an unexpected W-shaped structure in the reconstructed behavior of $w(z)$. It is interesting to note that this W-shaped feature emerges at a redshift around $z \simeq 0.51$, which is similar to our characteristic redshift $z_2 \sim 0.512$, where we observe $5\sigma$ deviation from the Planck $\Lambda$CDM model. \\

\begin{acknowledgments}
The authors acknowledge comments and suggestions from Eoin \'O Colg\'ain, Eleonora Di Valentino, Shahin Sheikh-Jabbari, Sunny Vagnozzi and Savvas Nesseris. PM acknowledges funding from the Anusandhan National Research Foundation (ANRF), Govt of India, under the National Post-Doctoral Fellowship (File no. PDF/2023/001986). AAS acknowledges the funding from ANRF, Govt of India, under the research grant no. CRG/2023/003984. We acknowledge the use of the HPC facility, Pegasus, at IUCAA, Pune, India. 
\end{acknowledgments}

\appendix



\bibliography{references}

\end{document}